\documentstyle[twoside,fleqn,espcrc2]{article}

\newcommand{\AmS}{{\protect\the\textfont2
  A\kern-.1667em\lower.5ex\hbox{M}\kern-.125emS}}

\title{\hfill{\normalsize IFUP-TH-52/96,~ 
              \texttt{hep-ph/9608415}} \\
\vspace{.2in}
$R$-parity-Violating Supersymmetric Yukawa Couplings: A Mini-review}

\author{Gautam Bhattacharyya
\address{Dipartimento di Fisica, Universit\`{a} di Pisa
and INFN, Sezione di Pisa, I-56126 Pisa, Italy}%
        \thanks{Invited talk given at the 4th International Conference
                on Supersymmetry (SUSY 96), College Park, Maryland, 
                May 29 -- June 1, 1996.}}
       
\begin{document}

\begin{abstract}
I review the bounds on the $R$-parity-violating supersymmetric Yukawa 
couplings from the considerations of  
proton stability, $n$--$\bar{n}$ oscillation, $\nu_e$-Majorana mass,
neutrino-less double $\beta$ decay, charged-current universality, 
$e$--$\mu$--$\tau$ universality, $\nu_\mu$--$e$ scattering, atomic 
parity violation, $\nu_\mu$ deep-inelastic scattering,  
$K^{+}$-decays, $\tau$-decays, $D$-decays and from 
the precision LEP electroweak observables. I also mention about the
sparticle bounds at colliders when 
the assumption of $R$-parity-conservation is relaxed. 
Finally, I mention how $R$-parity-violating models have been invoked in 
an attempt to explain the reported excess in ALEPH 4-jet events. 
\end{abstract}

\maketitle

\section{INTRODUCTION}

`$R$-parity' in supersymmetry (SUSY) refers to a discrete symmetry which
follows from the conservation of lepton-number ($L$) and baryon-number
($B$) \cite{rpar}. It is defined as
$R = (-1)^{(3B+L+2S)}$, where $S$ is the intrinsic spin of the
field. $R$ is $+1$ for all standard model (SM) particles and
$-1$ for all super-particles.
However, $B$- and $L$-conservations are not ensured by gauge
invariance and hence there is {\it a priori} no reason to set these 
couplings to zero. It is, therefore, a phenomenological 
exercise to constrain these couplings from observed and unobserved 
phenomena in nature.  
Minimal SUSY requires the presence of two Higgs superfield doublets and 
one of their gauge quantum numbers are the same as those of the
$SU(2)$-doublet lepton superfield. So, in the Yukawa superpotential, 
the latter can replace the former, if one sacrifices the
assumption of $L$-conservation. If one sacrifices the assumption 
of $B$-conservation as well, no theoretical
consideration prevents one to construct a 
Yukawa interaction involving three $SU(2)$-singlet quark superfields.
These lead to {\it explicit} breaking of $R$-conserving interactions,
which are parametrized in the superpotential (ignoring the bilinear
$\mu L_i H_2$ term) as
\begin{eqnarray}
    {\cal W}_{\not R} & = & \lambda_{ijk} L_i L_j E^c_k
                        +  \lambda'_{ijk} L_i Q_j D^c_k \nonumber \\
                      &  + & \lambda''_{ijk} U^c_i D^c_j D^c_k \ ,
\label{R-parity}
\end{eqnarray}
where $L_i$ and $Q_i $ are the $SU(2)$-doublet lepton and quark
superfields and $E^c_i, U^c_i, D^c_i$
are the singlet  superfields; $\lambda_{ijk}$ is antisymmetric
under the interchange of the first two $SU(2)$ indices,
while $\lambda''_{ijk}$ is antisymmetric under the interchange of
the last two. Thus, in total, there are 27 $\lambda'$-type and
9 each of $\lambda$- and $\lambda''$-type couplings, thereby adding
45 extra parameters in the minimal SUSY. 
 
\subsection{Cosmological implications}
The requirement that GUT-scale baryogenesis does not get washed out
imposes $\lambda''\ll 10^{-7}$ \cite{cosm}; however, these bounds are
model dependent and can be evaded \cite{dr}.
The $\lambda'$ couplings alone cannot wash out the initial baryon
asymmetry. But, they can do so in association
with a $B$-violating but $(B-L)$ conserving interaction, such as
sphaleron-induced non-perturbative transitions. The latter 
processes conserve ${1\over3}B - L_i$ for each lepton generation,
and hence the conservation of any one lepton generation number is enough 
to retain the initial baryon asymmetry. We, therefore, assume 
that the $\lambda'$-type couplings involving any particular lepton 
family are smaller than $\sim 10^{-7}$ to avoid any cosmological bound 
on the remaining of them.

\section{LOW ENERGY PHENOMENOLOGY}

\subsection{Proton stability}
Non-observation of proton decay places very strong bounds on the 
simultaneous presence of both $L$- and $B$-violating couplings; 
generically $\lambda' \lambda'' \le 10^{-24}$. 
Specific cases have been considered in refs. \cite{sher,vissani}. 
Ref.\cite{vissani} sets an upper limit of $10^{-9}$ ($10^{-11}$) 
for any product combination of $\lambda'$ and $\lambda''$ in the absence 
(presence) of squark flavour mixing. 

\subsection{$n$--$\bar{n}$ oscillation}
The contributions of the $\lambda''_{121}$- and 
$\lambda''_{131}$-induced interactions to $n$--$\bar{n}$ oscillation
proceed through the process $(udd\rightarrow \bar{\tilde{d_i}}d 
\rightarrow \tilde{g}\rightarrow \tilde{d_i}\bar{d}\rightarrow 
\bar{u}\bar{d}\bar{d})$. In ref.\cite{zwir}, the intergenerational 
mixing was not handled with sufficient care. In the updated analysis
\cite{goity}, the constraint on $\lambda''_{131}$ has been estimated 
to be $\le 10^{-4}-10^{-5}$ for $\tilde{m} = 100$ GeV, while that on 
$\lambda''_{121}$ is shown to be weaker (diluted by a relative factor of
$m_s^2/m_b^2$). It has, however, been shown in the same paper
\cite{goity} that 
the best constraint on $\lambda''_{121}$ comes from the consideration of 
double nucleon decay into two kaons and the bound is estimated to be 
$\le 10^{-6}-10^{-7}$.    

\subsection{$\nu_e$-Majorana mass}
$\lambda$- and $\lambda'$-type couplings can induce a Majorana mass 
of $\nu_e$ by self-energy type diagrams. An approximate expression for
the induced $\nu_e$-Majorana mass, for a generic coupling $\lambda$, is
\begin{equation}
\delta m_{\nu_e} \sim {\lambda^2 \over {8\pi^2}} {1\over {\tilde{m}^2}}
                                    M_{\rm SUSY} m^2.
\end{equation}
Assuming $M_{\rm SUSY}=\tilde{m}$, the $\lambda_{133}$-induced interaction 
with $\tau\tilde{\tau}$ loops yields the constraint ($1\sigma$)  
$\lambda_{133} \le 3 \times 10^{-3}$ for $m_{\tilde{\tau}} = 100$ 
GeV \cite{dimo}. On the other hand, the  
$\lambda'_{133}$-induced diagrams with $b\tilde{b}$ loops leads to  
$\lambda'_{133} \le 10^{-3}$ for $m_{\tilde{b}} = 100$ GeV \cite{grt}. 

\subsection{Neutrinoless double beta decay}
It is known for a long time that neutrinoless double beta decay
($\beta\beta_{0\nu}$) is a sensitive probe of lepton-number-violating
processes. In $R$-parity-violating scenario, the process 
$dd \rightarrow uue^{-}e^{-}$ is mediated by $\tilde{e}$ and 
$\tilde{\gamma}$ or by $\tilde{q}$ and $\tilde{g}$, yielding  
$\lambda'_{111} \le 10^{-4}$ \cite{rabi_db,klap}. 
Recently, a new bound on the  
product coupling $\lambda'_{113} \lambda'_{131} \le 3 \times 10^{-8}$
has been placed from the consideration of the diagrams involving the 
exchange of one $W$ boson and one scalar boson \cite{babu_db}. 

\subsection{Charged-current universality}
Universality of the lepton and quark couplings to the $W$-boson is 
violated by the presence of $\lambda$- and $\lambda'$-type couplings.
The scalar-mediated new interactions have the same $(V-A)
\otimes (V-A)$ structure as the $W$-exchanged diagram. The experimental 
value of $V_{ud}$ is related to $V_{ud}^{\rm SM}$ by 
\begin{equation}
|V_{ud}^{\rm exp}|^2 \simeq |V_{ud}^{\rm SM}|^2 \left[1+{{2 r'_{11k}
(\tilde{d}_R^k)}\over {V_{ud}}} - 2 r_{12k}(\tilde{e}_R^k)\right],
\end{equation}
where, 
\begin{equation}
r_{ijk}(\tilde{l}) = (M_W^2/g^2)(\lambda_{ijk}^2/m_{\tilde{l}}^2). 
\end{equation}
$r'_{ijk}$ is defined using $\lambda'_{ijk}$
analogously as $r_{ijk}$. Assuming the presence of 
only one $R$-parity-violating coupling at a time, one obtains, for a 
common $\tilde{m} = 100$ GeV, $\lambda_{12k} \le 0.04~(1\sigma)$ and 
$\lambda'_{11k} \le 0.03~(2\sigma)$, for each $k$ \cite{bgh}. 

\subsection{$e$--$\mu$--$\tau$ universality}
The ratio $R_{\pi} \equiv \Gamma(\pi \rightarrow e\nu)/
\Gamma(\pi \rightarrow \mu\nu)$, in the presence of $\lambda'$-type 
interaction takes the form 
\begin{equation}
R_\pi = R_\pi^{\rm SM} \left[1+ {2\over V_{ud}}\left\{r'_{11k}
(\tilde{d}_R^k) - r'_{21k}(\tilde{d}_R^k)\right\}\right].
\end{equation}
A comparison with experimental results yields, for a common mass 
$\tilde{m} = 100$ GeV and at $1\sigma$, 
$\lambda'_{11k} \le 0.05$ and $\lambda'_{21k} \le 0.09$, for each $k$, 
assuming only one coupling at a time \cite{bgh}.  

Similarly, from the consideration of $R_\tau \equiv \Gamma(\tau 
\rightarrow e\nu\bar{\nu})/ \Gamma(\tau \rightarrow \mu\nu\bar{\nu})$, 
one obtains, $\lambda_{13k} \le 0.10$ and $\lambda_{23k} \le 0.12$, for 
each $k$, at $1\sigma$ and for $\tilde{m} = 100$ GeV \cite{bgh}.  

\subsection{$\nu_\mu$--$e$ scattering}
The neutrino-electron scattering cross section at low energies are 
given by 
\begin{eqnarray}
\sigma(\nu_\mu e) & = & {G_F^2 s\over{\pi}}(g_L^2 + {1\over 3} g_R^2),
                                         \nonumber \\
\sigma(\bar{\nu}_\mu e) & = & {G_F^2 s\over{\pi}}
({1\over 3}g_L^2 +  g_R^2); 
\end{eqnarray}
where in the presence of $R$-parity-violating interactions 
$(x_W \equiv \sin^2\theta_W)$
\begin{eqnarray}
g_L & = & x_W - {1\over 2} - ({1\over 2} + x_W) r_{12k}(\tilde{e}_R^k),
                                     \nonumber \\
g_R & = & x_W + r_{121}(\tilde{e}_L^1) + r_{231}(\tilde{e}_L^3) \\
    & - &  x_W r_{12k}(\tilde{e}_R^k). \nonumber
\end{eqnarray}
The derived constraints (at $1\sigma$) are $\lambda_{12k} \le 0.34$, 
$\lambda_{121} \le 0.29$ and $\lambda_{231} \le 0.26$ for $\tilde{m} 
= 100$ GeV \cite{bgh}.

\subsection{Atomic parity violation}
The parity-violating part of the Hamiltonian of the electron-hadron 
interaction is 
\begin{equation}
H = {G_F\over{\sqrt{2}}} \left(C_{1i} \bar{e}\gamma_\mu\gamma_5 e
\bar{q}_i\gamma_\mu q_i + C_{2i} \bar{e}\gamma_\mu e
\bar{q}_i\gamma_\mu \gamma_5 q_i\right), 
\end{equation}
where, $i$ runs over the $u$- and $d$-quarks. 
For the definitions of the $C_i$'s in the SM , see any 
Review of Particle Properties (e.g., ref.\cite{pdg}).  
The $R$-parity violating contributions are ($\Delta C \equiv 
C - C^{\rm SM}$), 
\begin{eqnarray}
\Delta C_1^u & = &  - r'_{11k}(\tilde{d}_R^k) + ({1\over 2} 
             - {4\over 3} x_W) r_{12k}(\tilde{e}_R^k), \nonumber \\
\Delta C_2^u & = & - r'_{11k}(\tilde{d}_R^k) + ({1\over 2} 
             -           2 x_W) r_{12k}(\tilde{e}_R^k), \nonumber \\
\Delta C_1^d & = &  r'_{1j1}(\tilde{q}_L^j) - ({1\over 2} 
             - {2\over 3} x_W) r_{12k}(\tilde{e}_R^k),  \\
\Delta C_2^d & = & - r'_{1j1}(\tilde{q}_L^j) - ({1\over 2} 
             -           2 x_W) r_{12k}(\tilde{e}_R^k). \nonumber
\end{eqnarray}
Including the effects of radiative corrections, the $1\sigma$ bounds are 
$\lambda'_{11k} \le 0.30, \lambda'_{1j1} \le 0.26$ for $\tilde{m} = 100$
GeV \cite{bgh}. Bounds on $\lambda_{12k}$ are much weaker than those 
obtained from charged-current universality.    

\subsection{$\nu_\mu$ deep-inelastic scattering}
The left- and the right-handed couplings of the $d$-quark in neutrino 
interactions are modified by the $R$-parity-violating couplings as 
\begin{eqnarray} 
g_L^d & = &(-{1\over 2} + {1\over 3} x_W)(1-r_{12k} (\tilde{e}_R^k))
 - r'_{21k} (\tilde{d}_R^k), \nonumber \\
g_R^d & = & {1\over 3} x_W + r'_{2j1} (\tilde{d}_L^j)
            -{1\over 3} x_W r_{12k} (\tilde{e}_R^k). 
\end{eqnarray}
The derived limits, for $\tilde{m} = 100$ GeV, are 
$\lambda'_{21k} \le 0.11$ ($1\sigma$) and 
$\lambda'_{2j1} \le 0.22$ ($2\sigma$) \cite{bgh}. 

\subsection{$K^{+}$-decays}
Consideration of only one non-zero $R$-parity-violating coupling with 
indices related to the weak basis of fermions, automatically generates  
more than one non-zero coupling with different flavour structure in 
the mass basis. Consequently, flavour-changing-neutral-current processes
are naturally induced. The Lagrangian governing $K^{+} \rightarrow 
\pi^+ \nu\bar{\nu}$ is given by
\begin{equation}
L = - {{\lambda'^2_{ijk}} \over{2 m_{\tilde{d}_R^k}^2}} V_{j1} V^*_{j2}
(\bar{s}_L\gamma^\mu d_L)(\bar{\nu}_{Li} \gamma_\mu \nu_{Li}), 
\end{equation}
where V is the CKM matrix. The SM contribution is an order of magnitude 
lower than the experimental limit. Assuming 
that the new interaction dominates,
one obtains, from the ratio of the $\Gamma(K^+ \rightarrow \pi^+\nu_i
\bar{\nu}_i)$ to $\Gamma(K^+ \rightarrow \pi^0\nu\bar{e})$, the constraint
 $\lambda'_{ijk} \le 0.012$ (90\% CL), for $m_{\tilde{d}_R^k} = 100$ GeV
and for $j = 1$ and $2$ \cite{agashe}. 

\subsection{$\tau$-decays}
The decay $\tau^{-} \rightarrow \bar{u}d\nu_\tau$ proceeds in the SM by 
a tree-level $W$-exchanged graph. The scalar-exchanged graph induced
by $\lambda'_{31k}$ can be written in the same $(V-A)\otimes (V-A)$
form by a Fierz rearrangement. Using the experimental input  
\begin{eqnarray}
   Br(\tau^- \rightarrow \pi^- \nu_\tau) & = & 0.117 \pm 0.004, \\
   f_\pi &=& (130.7 \pm 0.1 \pm 0.36) \; {\rm MeV} \nonumber.
   \label{fpi}
\end{eqnarray}
one obtains $\lambda'_{31k}  \le  0.16$ ($1\sigma$) for  
$m_{\tilde{d}_R^k} = 100$ GeV \cite{dtau}. 

\subsection{$D$-decays}  
The tree-level process $c \rightarrow s e^+ \nu_e$ is mediated 
by a $W$ exchange in the SM and by a scalar boson exchange in 
$\lambda'$-induced interaction. By a Fierz transformation it is 
possible to write the latter in the same $(V-A)\otimes (V-A)$ form
as the former. Using the experimental input \cite{pdg}:
\begin{equation}
 \frac{ Br(D^+ \rightarrow \bar{K}^{0 \ast} \mu^+ \nu_\mu ) }
        { Br(D^+ \rightarrow \bar{K}^{0 \ast} e^+ \nu_e ) }
     =  0.94 \pm 0.16, 
\end{equation}
one obtains, at $1\sigma$, $\lambda'_{12k} \le  0.29$ and 
$\lambda'_{22k} \le  0.18$, for $m_{\tilde{q}} = 100$ GeV   
\cite{dtau}. The form factors associated
with the hadronic matrix elements cancel in the ratios, thus
making the prediction free from the large theoretical
uncertainties associated with those matrix elements.

\begin{table*}[hbt]
\setlength{\tabcolsep}{1.5pc}
\newlength{\digitwidth} \settowidth{\digitwidth}{\rm 0}
\catcode`?=\active \def?{\kern\digitwidth}
\caption{The most stringent constraints on $R$-parity-violating couplings 
         for $\tilde{m} = 100$ GeV ($m = 1,2$; $n,l = 1,2,3$). The 
         remaining four $\lambda''$-couplings, which are not listed  
         below, are constrained only from the requirement of perturbative 
         unitarity ($\le$ 1.25) [5,20]. Note, in the last row $n\ne l$.}
\label{tab:couplings}
\begin{tabular*}{\textwidth}{@{}l@{\extracolsep{\fill}}rrrrr}
\hline
\hline 
$(12n)$   & 0.04~(a)  & $(13n)$ & 0.10~(b)   & $(23n)$  & 0.09~(c) \\
\hline
$(1mn)'$  & 0.012~(d) & $(2mn)'$ & 0.012~(d) & $(3mn)'$ & 0.012~(d) \\
$(131)'$  & 0.26~(e)  & $(231)'$ & 0.22~ (g) & $(331)'$ & 0.26~(h) \\
$(132)'$  & 0.4~(f)   & $(232)'$ & 0.4~  (f) & $(332)'$ & 0.26~(h) \\
$(133)'$  & 0.001~(i) & $(233)'$ & 0.4~  (f) & $(333)'$ & 0.26~(h) \\
\hline 
$(112)''$ & $\sim 10^{-6}$~(j) & 
$(113)''$ & $\sim 10^{-4}$~(k) & $(3nl)''$ & 0.97~(l) \\
\hline
\hline
\multicolumn{5}{@{}p{120mm}}{(a): Charged-current universality 
                                    ($1\sigma$) \cite{bgh},
                            ~(b): $\Gamma(\tau \rightarrow e\nu\bar{\nu})/ 
                                  \Gamma(\tau \rightarrow \mu\nu\bar{\nu})$
                                   ($1\sigma$) \cite{bgh},
                            ~(c): $\Gamma(\tau\rightarrow \mu\nu\bar{\nu})/ 
                                  \Gamma(\mu \rightarrow e\nu\bar{\nu})$
                                   ($1\sigma$) \cite{bgh},      
                            ~(d): $K^+$-decay (90\% CL) \cite{agashe},
                            ~(e): Atomic parity violation and $eD$
                                   asymmetry ($1\sigma$) \cite{bgh}, 
                            ~(f): $t$-decay (2$\sigma$) \cite{agashe}, 
                            ~(g): $\nu_\mu$ deep-inelastic scattering
                                   ($2\sigma$) \cite{bgh}, 
                            ~(h): $Z$ decay width ($1\sigma$) \cite{bes},
                            ~(i): $\nu_e$ mass ($1\sigma$) \cite{grt},
                            ~(j): double nucleon decay ($1\sigma$)
                                   \cite{goity}, 
                            ~(k): $n$--$\bar{n}$ oscillation ($1\sigma$)
                                   \cite{goity},
                            ~(l): $Z$ decay width ($1\sigma$) \cite{bcs}. }
\end{tabular*}
\end{table*}

\section{LEP PRECISION MEASUREMENTS}

The partial decay widths ($\Gamma_i$) of the $Z$ boson into light 
fermions receive sizable triangle-loop corrections when heavy chiral 
fermions float inside the loops. 
The $\lambda'_{ijk}$-induced vertex corrections
involve new triangle diagrams contributing to $\Gamma_l$
with $Z, l^+$ and $l^-$ as external lines  where $i =$
lepton, $j =$ quark, $k =$ squark indices or $i =$ lepton, $j =$
squark, $k =$ quark indices.
Such couplings also add corrections to $\Gamma_{\rm had}$ through 
triangle diagrams where the external
lines are $Z, q$ and $\bar{q}$ in a situation where, for example,
$i =$ slepton, $j =$ quark (squark) and $k =$ squark (quark).
Since the heaviness of the chiral fermion in the loop is the crucial
factor in determining the size of the new contributions, only
$\lambda'_{i3k}$-type couplings involving internal top quark lines
are constrained significantly by such processes \cite{bes}.
Similarly, the $\lambda''$-induced corrections to the decay vertices
$Z\rightarrow q\bar{q}$ also add sizable corrections to the  
hadronic partial widths \cite{bcs}. 
Consequently, for $\tilde{m} = 100$ GeV and at $1\sigma$,
the following bounds emerge 
($R_l = \Gamma_{\rm had}/\Gamma_l$):\footnote{While
extracting limits on $\lambda''$, leptonic universality in $R_l$ 
is assumed since $\lambda''$-Yukawa's do not involve any leptonic flavour.}
\begin{eqnarray}
\lambda'_{13k} & \le & 0.51 \leftarrow
R_e^{\rm exp} = 20.850 \pm 0.067, \nonumber \\
\lambda'_{23k} & \le & 0.44 \leftarrow
R_\mu^{\rm exp} = 20.824 \pm 0.059, \nonumber \\
\lambda'_{33k} & \le & 0.26 \leftarrow
R_\tau^{\rm exp} = 20.749 \pm 0.070, \\
\lambda''_{3jk} & \le & 0.97 \leftarrow
R_l^{\rm exp} = 20.795 \pm 0.040. \nonumber
\end{eqnarray}
The above experimental input are collected from the LEP
Electroweak Working Group report \cite{lep}.

\section{DIRECT SEARCHES AT COLLIDERS}

\subsection{LEP1}
In the $R$-parity-violating scenario, the LSP is unstable. 
The OPAL Collaboration at LEP \cite{opal} have assumed the photinos 
to be the LSP's decaying {\it via} a $\lambda_{123}$-type coupling. 
They excluded at 95\% C.L.
$m_{\tilde{\gamma}} =$ 4--43 GeV for $m_{\tilde{e}_L} < 42$ GeV,
and $m_{\tilde{\gamma}} = $ 7--30 GeV for $m_{\tilde{e}_L} < 100$ GeV.

The ALEPH Collaboration at LEP \cite{aleph}, dealing with a more
general $\lambda$-type coupling and considering a general LSP
rather than a pure photino, have updated the above exclusion zone
and have also reported their negative results on other
supersymmetric particles up to their kinematic limit ($<M_Z/2$).

A lighter photino ($\sim$ 2--3 GeV) in conjunction with a 
$R$-parity-violating coupling provides a new semileptonic $B$-decay 
mode ($b \rightarrow ce\tilde{\gamma}$). Arranging such that the 
photino does not decay within the detector, the above channel adds 
incoherently to the standard semileptonic decay mode. 
However, the new mode, owing to the  
massive nature of the photino, arranges a different kinematic 
configuration compared to the standard channel where neutrino 
carries the missing energy. A kinematic exploration 
of the above has been carried out in the context of LEP and CLEO 
\cite{gbar}\footnote{Light photino with $R$-parity-violation 
has been employed \cite{subir} to resolve the KARMEN anomaly.}.      

\subsection{LEP2}
The $\tau$-number-violating operators were studied in the context of 
LEP2 in ref. \cite{grt}. Like-sign di-tau events accompanied by jets
without any missing $E_T$ were predicted as the most spectacular signals
of such interactions. 

Indirect effects of $R$-parity-violating couplings
through deviations in the angular distributions of $e^+e^- \rightarrow 
f\bar{f}$ due to the induced sfermion-exchanged diagrams have been 
studied \cite{bhondo} at LEP2 energies.  

\subsection{Fermilab Tevatron}
The impact of the $\lambda'$-type couplings in $t$-quark decay at the 
Tevatron have been analysed in ref. \cite{agashe}. One of the consequences 
is the following: In the SM, the dominant decay mode is 
$t \rightarrow b W$. The $\lambda'_{i3k}$-type couplings will induce 
$t_L \rightarrow \tilde{l}_i^+ d_{Rk}$ (if kinematically allowed), 
followed by $\tilde{l}_i^+ \rightarrow l^+ \tilde{\chi}^0$ (100\%) and 
$\tilde{\chi}^0 \rightarrow (\nu_i + b + \bar{d}_k, 
~ \bar{\nu}_i + \bar{b} + d_k)$
leading to final states with at one lepton, at least one $b$-quark and 
missing $E_T$. The characteristic features of this decay channel are 
that it spoils the lepton universality and for $k = 3$ produces 
additional $b$-quark events.

Strategies of setting squark and gluino mass limits from multilepton 
final states in the absence of $R$-parity-conservation have been discussed
in ref. \cite{dp}.  

\section{ALEPH 4-JET ANOMALY}
On the basis of the LEP 1.5 run at $\sqrt{s} = 130$ -- 136 GeV, the 
ALEPH Collaboration have reported \cite{aleph4j} an excess number of 
events in $e^+e^- \rightarrow$ 4 jets channel. They observed 16 events
where the SM predicts 8.6. The excess 9 events have a 4-jet invariant
mass $\Sigma M = 105$ GeV. There have been a few attempts to explain 
this anomaly by invoking the $R$-parity-violating couplings: 

\begin{enumerate}
\item  Refs.\cite{bkp,peccei} consider the pair production of sfermions
by gauge interactions and their subsequent decays by $L$-violating 
(sneutrino decays \cite{bkp})- or $B$-violating (squark decays 
\cite{peccei})- couplings to quarks. Thus, although, notionally these lead 
to 4-jet final states, owing to small sfermion production cross section, 
enough number of events do not survive after the imposition of the ALEPH 
cuts. 

\item Ref.\cite{teni} considers, as the most optimistic option, the 
pair production of charginos ($e^+e^- \rightarrow \tilde{\chi}^+   
\tilde{\chi}^-$), followed by $\tilde{\chi}^+ \rightarrow \tilde{\chi}_1^0
({\rm LSP}) + W^{+*}$, and finally the $\lambda''$-induced decay  
$\tilde{\chi}_1^0 \rightarrow u_id_jd_k$ (and similar combinations)
{\it via} virtual squark states. 
If the off-shell $W^{*}$'s decay hadronically,
then there are 10 jets in the final states, which are required to merge
into 4 somewhat fat jets. This has been claimed as a viable option. 
In the case of leptonic decay of one $W^{*}$, the final state leptons
can escape detection by lying within the jets and after jet-merging a few 
4-jet events still survive.  

\item Ref.\cite{herbi} interpretes the observed excess in 4-jet events 
as $e^+e^- \rightarrow \tilde{\chi}^+ \tilde{\chi}^- \rightarrow 
d_j\bar{d}_k\bar{d}_jd_k\tau^+\tau^-$, where the chargino decays are 
induced by $\lambda'_{3jk}$-couplings. Thus, the final states contain 
4 jets and 2 soft $\tau$'s which are experimentally reconstructed as 
4 jets.       
   
\item Ref.\cite{ccp} considers the pair production of charginos 
and finds the best solution to be $\tilde{\chi}_1^- \rightarrow
\bar{\tilde{t}}_1 b \rightarrow dsb$ (the $\tilde{t}_1$ decay is induced by 
$\lambda''$), with the extremely soft $b$ evading detection as a result
of the kinematic choice: $m_{\tilde{\chi}_1} \simeq 60$ GeV and 
$m_{\tilde{t}} \simeq 52$ GeV. 
\end{enumerate}

The main message that can be read from the above analyses is that the 
pair production cross section of charginos are sigificantly higher than 
those of the sfermions (and also higher than neutralino pair production 
cross section) and, therefore, even after paying the price of losing 
events while imposing the kinematic cuts during cascades
following the decays of the charginos, required number of 
events still manage to survive resembling the 4-jet excess. But, most 
importantly, before speculating further, one should wait and see whether 
these anomalous events stand the test of time!!    

\section{CONCLUSION} 
In this talk, I have reviewed the existing bounds on the 
$R$-parity-violating couplings from low energy data and from 
LEP1 data. While the low energy data tend to constrain more the couplings
involving the lighter generations, the LEP data are rather sensitive to 
couplings involving the third generation. The implications of 
$R$-parity-violation on direct searches at colliders are also 
mentioned. The excess 4-jet events at the LEP 1.5 run reported by 
the ALEPH Collaboration could find a natural explanation in the 
$R$-parity-violating atmosphere. 

The effects of $R$-parity-violation in the context of GUT were 
discussed by F. Vissani and the RG-evolutions of those couplings with an 
emphasis on the fixed point solutions were discussed by V. Barger 
in this Conference.

I thank D. Choudhury, J. Ellis, A. Raychaudhuri and K. Sridhar for 
stimulating collaborations on various aspects of $R$-parity-violation.
I also thank the Organizers  of SUSY 96 for invitation.

\end{document}